\documentclass[10pt,conference]{IEEEtran}

\IEEEoverridecommandlockouts

\usepackage[cmex10]{amsmath}
\usepackage{amssymb}

\usepackage{algorithm, algpseudocode}

\usepackage{graphicx}
\usepackage[capitalise]{cleveref}

\ifCLASSOPTIONcompsoc	\usepackage[caption=false,font=normalsize,labelfont=sf,textfont=sf]{subfig}
\else \usepackage[caption=false,font=footnotesize]{subfig}
\fi

\ifCLASSOPTIONcompsoc \usepackage[nocompress]{cite}
\else                 \usepackage{cite}
\fi

\begin{document}

\title{Joint Channel Bandwidth Assignment and Relay Positioning for Predictive Flying Networks}

\author{
    \IEEEauthorblockN{
        Ruben Queiros\IEEEauthorrefmark{1},
        Megumi Kaneko\IEEEauthorrefmark{2},
        Helder Fontes\IEEEauthorrefmark{1},
        Rui Campos\IEEEauthorrefmark{1}
    }
    \IEEEauthorblockA{
        \IEEEauthorrefmark{1}INESC TEC and Faculdade de Engenharia, Universidade do Porto, Portugal \\
        \{ruben.m.queiros, helder.m.fontes, rui.l.campos\}@inesctec.pt, \\
        \IEEEauthorrefmark{2}National Institute of Informatics, Tokyo Japan, megkaneko@nii.ac.jp
    }   
    \thanks{This work is co-financed by Component 5 - Capitalization and Business Innovation of core funding for Technology and Innovation Centres (CTI), integrated in the Resilience Dimension of the Recovery and Resilience Plan within the scope of the Recovery and Resilience Mechanism (MRR) of the European Union (EU), framed in the Next Generation EU, for the period 2021 - 2026, by National Funds through the Portuguese funding agency, FCT - Fundação para a Ciência e Tecnologia, under the PhD grant 2022.10093.BD, by the NII MoU grant between NII and INESC TEC, and by the Grants-in-Aid for Scientific Research no. 20H00592 and 22KK0156 from the Ministry of Education, Science, Sports, and Culture of Japan.
    }
    \vspace{-1cm}
}

\maketitle

\begin{abstract}
Flying Networks (FNs) have emerged as a promising solution to provide on-demand wireless connectivity when network coverage is insufficient or the communications infrastructure is compromised, such as in disaster management scenarios. Despite extensive research on Unmanned Aerial Vehicle (UAV) positioning and radio resource allocation, the challenge of ensuring reliable traffic relay through backhaul links in predictive FNs remains unexplored.

This work proposes Simulated Annealing for predictive FNs (SAFnet), an innovative algorithm that optimizes network performance under positioning constraints, limited bandwidth and minimum rate requirements. Our algorithm uniquely leverages prior knowledge of the first-tier node trajectories to assign bandwidth and dynamically adjust the position of the second-tier flying relay. Building upon Simulated Annealing, our approach enhances this well-known AI algorithm with penalty functions, achieving performance levels comparable to exhaustive search while significantly reducing computational complexity.

\end{abstract}

\begin{IEEEkeywords}
Flying Networks, Relay Positioning, Channel Assignment, Wireless Communications
\vspace{-0.2cm}
\end{IEEEkeywords}

\section{Introduction}

Natural and man-made disasters can rapidly bring down existing communications infrastructure, creating urgent demand for adaptable, resilient connectivity solutions. Flying networks (FNs), composed of Unmanned Aerial Vehicles equipped with communications hardware payloads, offer a scalable, cost-effective solution to extend and reinforce network coverage and capacity in such critical scenarios, enabling the deployment of intelligent emergency communications. Unlike fixed base stations, UAVs can be redeployed multiple times, and their inherent mobility enables rapid, on-the-fly adjustments to the network topology and positioning -- capabilities far beyond those of conventional Cell on Wheels solutions~\cite{cell-on-wheels}.

\begin{figure}
  \centering
  \includegraphics[width=.7\linewidth]{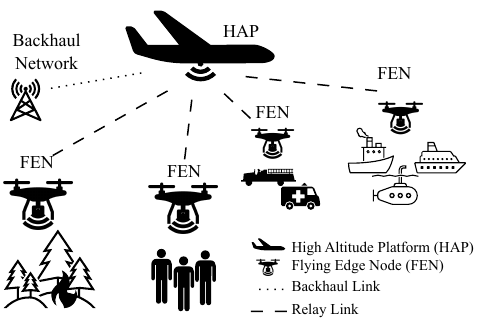}
\vspace{-0.4 cm}
  \caption{Two-tier predictive Flying Network.}
  \label{fig:concept}
\vspace{-0.8 cm}

\end{figure}

Despite the potential of FNs, key challenges persist, particularly in the optimal positioning and real-time allocation of radio resources. Existing research has extensively explored these aspects, yet it often overlooks the dynamics of predictive FNs, where UAVs follow predefined trajectories with specific Quality of Service (QoS) requirements~\cite{facom-survey}. We address this scenario, where the Flying Edge Nodes (FEN) have previously defined missions~\cite{respondrone} with known QoS requirements and predefined trajectories assigned by the mission commander. Figure~\ref{fig:concept} illustrates a predictive FN, where the FENs have their mission assigned, and the High Altitude Platform (HAP) has the role of ensuring the relay of communications to a backhaul network. This network is not directly communicating with FENs due to differences in wireless technology or communications range. In addition to HAP positioning, the management of limited wireless channels is crucial to ensure resource allocation that maximizes network performance while minimizing radio interference. Furthermore, low-complexity algorithms are needed to meet decision time requirements and reduce energy consumption, which are key aspects in FNs. 

In recent years, several solutions have been proposed to address the UAV optimal positioning. In~\cite{Almeida_2020,Coelho_2019} the authors address the joint optimization of the UAV placement and routing. However, they consider static default values for the channel bandwidth, and overlook the link to the backhaul network. In~\cite{rw2}, the authors maximize the number of connections in an FN that satisfy minimum QoS requirements, by optimizing radio resources, node trajectory and computing resources but the computational complexity of the proposed method does not scale with the number of nodes. Reference~\cite{rw3} proposes a low-complexity algorithm that maximizes the minimum throughput for the ground users served by the FN, considering the optimization of the UAV trajectories and transmission power. However, similarly to~\cite{rw2}, the authors do not consider the link to the backhaul network nor predefined FEN trajectories. A backhaul-aware solution has been proposed in~\cite{rw4}, but using branch-and-bound and discrete grid search to determine the optimal UAV position is computationally expensive, potentially preventing a timely computation of the solution. In~\cite{sabzehali2022}, the authors optimize the number and position of UAVs in an FN considering the backhaul link and using orthogonal frequency channels. Yet, they neglect the limited backhaul link capacity and channel bandwidth. In~\cite{2019globecom}, the authors optimize the UAV positioning and bandwidth allocation in single-tier FNs focusing on the ground users' links. They consider the backhaul link but assume the bandwidth can be arbitrarily fractioned, which is not valid for real communication systems.

In this paper, we propose an innovative algorithm designed to optimize network performance under predefined FEN trajectories, limited bandwidth of the wireless links, minimum rate requirements for the FENs, and limited backhaul link capacity. Our algorithm uniquely leverages prior knowledge of the FENs' trajectories to dynamically adjust the position of the HAP and assign bandwidth to the relay and backhaul links (cf. Figure~\ref{fig:concept}). By utilizing Simulated Annealing (SA), our approach incorporates penalty functions to adapt this well-known AI algorithm to the problem, achieving performance levels comparable to exhaustive search while significantly reducing complexity.

The contributions of this paper are three-fold:
\begin{enumerate}
    \item  We mathematically formulate the joint problem of wireless channel assignment and HAP positioning in predictive FNs, considering constraints such as channel bandwidth, backhaul capacity and minimum rate requirements of the FENs;
    \item We design an approach based on SA to solve this intricate problem,  by incorporating constraint-specific penalty functions to effectively manage these constraints while maintaining low computational complexity;
    \item We provide numerical results that demonstrate the benefits of the proposed approach in terms of weighted sum-rate and outage probabilities, outperforming state-of-the-art benchmarks.

    \end{enumerate}

\section{System Model} 

The two-tier FN considered in this work consists of a HAP $h$ assigned to serve a group of \emph{E} FENs, relaying their traffic to a Backhaul Network node $b$, as shown in Fig.~\ref{fig:concept}. Let $j \in \{1,...,E\}$ denote the generic FEN. The zone to be covered by the FN is defined as a cuboid with dimensions: $Z_x \times Z_y \times Z_z$. We discretize time \emph{T} into \emph{N} time periods evenly spaced with duration $\Delta T$, so that $N=\frac{T}{\Delta T}$. Let $T_n$ denote the $n$-th time period, where $n \in \{1,...,N\}$. We adopt a 3D coordinate system, and the positions of the flying nodes are assumed to be fixed during each time period $T_n$. Therefore, the location of FEN $j$ is given by $\ell_{j}(T_n) = (x_{j}(T_n), y_{j}(T_n), z_{j}(T_n))$. Similarly, the HAP location is given by $\ell_{h} = (x_{h}, y_{h}, z_{h})$ and the location of the Backhaul Network node by $\ell_{b} = (x_{b}, y_{b}, z_{b})$. The FENs have their mission predefined at the beginning of $T$, and their positions over $T$ are known and given by the set $\mathcal{L}$.

Let $\mathcal{C}$ denote the set of supported channel bandwidth values by the IEEE 802.11 standard, and let $B_{F}$ be the available bandwidth in the FN's operating frequency band. We define $\mathcal{B}=\{B_{ih}\}$ with $i \in \{1,...,E,b\}$ as the set of orthogonal channel bandwidths assigned to each of the communication links, with $B_{bh} \in \mathcal{C}$ being the channel bandwidth allocated to the link between backhaul $b$ and HAP $h$. The system sets $\mathcal{B}$ at the beginning of time $T$, and it remains fixed over all $N$ periods. 

In FNs, the wireless channel is typically characterized by an obstacle-free strong Line of Sight component~\cite{channel-uav-survey}. Therefore, we model the large scale effects, i.e., the path loss $L_{jh}(T_n)$ of the channel  between $j$ and $h$ during time period $T_n$, using the Free Space Path Loss model as follows:
\begin{equation}
    L_{jh}(T_n)=\left(\frac{\lambda}{4\pi d_{jh}(T_n)}\right)^2,
    \label{eq:fspl}
\end{equation}
where $d_{jh}(T_n)=||\ell_j(T_n) - \ell_h(T_n)||$ is the Euclidean distance between $j$ and $h$, and $\lambda$ is the wavelength. The received SNR at $h$ from a transmitting node $j$, $\Gamma_{jh}(T_n)$, is then given by: \begin{equation}
    \Gamma_{jh}(T_n)=\frac{P^{T}_j L_{jh}(T_n)}{B_{jh}N_0} ,
    \label{eq:SNR}
\end{equation}
where $P^T_j$ is the fixed transmit power and $N_0$ is the power spectral density of the Additive White Gaussian Noise. Each link uses an orthogonal channel and there is no interference between nodes\footnote{Allocation with channel reuse, causing interference, will be investigated in the extended work.}. Finally, the link capacity is defined by the Shannon-Hartley theorem:
\begin{equation}
    \tau_{jh}(T_n) = B_{jh} \log_2(1+\Gamma_{jh}(T_n)),
    \label{eq:shannon}
\end{equation}
where $\tau_{jh}(T_n)$ is the capacity between FEN $j$ and HAP $h$ in bit/s during time period $T_n$.

\section{Problem Formulation}

By taking advantage of prior knowledge of the FENs' trajectories and rate requirements, the problem to be solved is the joint HAP positioning and orthogonal channel assignment for each link of the FN. The problem is mathematically formulated as follows:
\vspace{-0.1cm}
\begin{align}
\label{eq:problem1}
& \max_{\ell_h, \mathcal{B}}U(\ell_h, \mathcal{B})= \frac{1}{N}\sum_{n=1}^N \sum_{j=1}^E w_{j} \tau_{jh}(T_n) \\
\label{eq:c-location}
& \text{s.t.}~(0,0,0) \le (x_h,y_h,z_h) \le (Z_x,Z_y,Z_z),\\
\label{eq:c-samepos}
& d_{jh}(T_n) \geq d^{\mathrm{min}}, \forall j \in \{1,...,E\}, \forall n \in \{1,...,N\},   \\
\label{eq:c-channel}
& \sum_{B_{ih}\in \mathcal{B}} B_{ih} \leq B_{F}, \forall \mathcal{B} \in \mathcal{C}^{E+1}, \\
\label{eq:c-bkh}
& \sum_{j=1}^E\tau_{jh}(T_n) \leq \tau_{bh}(T_n), \forall n \in \{1,...,N\}, \\
\label{eq:c-min}
& \tau_{jh}(T_n) \geq \tau^{\mathrm{min}}_{jh}, \forall j \in \{1,...,E\}, \forall n \in \{1,...,N\},
\end{align}
where $w_{j}$ is the normalized weight assigned to FEN $j$. Eq.~(\ref{eq:c-location}) and Eq.~(\ref{eq:c-samepos}) limit the possible HAP locations to be within the network coverage boundaries and with a safe distance $d^{\mathrm{min}}$ from other moving FENs. Eq.~(\ref{eq:c-channel}) ensures that the total allocated channel bandwidth does not exceed the available bandwidth. Eq.~(\ref{eq:c-min}) defines $\tau^{\mathrm{min}}_{jh}$ as a minimum rate requirement for each FEN $j$ -- HAP link. Finally, Eq.~(\ref{eq:c-bkh}) ensures that the sum-rate over all FEN-HAP links is equal or below the backhaul link capacity.

This is a Mixed Integer Non Linear Problem (MINLP), which is intricate to solve. In challenging scenarios such as those involving a large number of FENs or high capacity requirements, it may not be possible to satisfy all constraints simultaneously. However, in real-world deployments, finding the best possible solution is still essential.

\section{Benchmark and Proposed Solutions}
\label{sec:refmethods}

This section first presents the benchmark methods to solve Problem~(\ref{eq:problem1})~$\sim$~(\ref{eq:c-min}), namely the conventional heuristic method, and the conventional SA method adapted for the problem at hand. Finally, the proposed method is detailed.

\subsection{Conventional Heuristic Method}

The Conventional Heuristic (Conv. H) method is the baseline solution described in Alg.~\ref{alg:heur}. The channel bandwidth is proportionally assigned to each node, considering their minimum rate requirements, and the discrete set of supported channel bandwidth values (Lines 3, 6). For the gateway position, it is common to assume the \textit{weighted centroid} of the cluster of UAVs~\cite{centroid}, considering the FENs trajectory knowledge $\mathcal{L}$ and minimum rate requirements $\tau^{\mathrm{min}}$ (Line 9). The output of Conv. H is used as an initial reasonable solution for both SA-based methods, as detailed next.

\begin{algorithm}
\caption{Conventional Heuristic Method}
\begin{algorithmic}[1]
\State \textbf{Input:}
$\mathcal{L}, \tau^{\mathrm{min}}, B_F $ \textbf{ Output:}
$\ell_h, \mathcal{B}$
\State {Initialize:} 
$\tau_{bh} = \sum \tau_{jh}^{\mathrm{min}}$, $\tau_{\mathrm{total}}=\tau_{bh} + \sum \tau_{jh}^{\mathrm{min}}$ 

\State $B_b = B_F \frac{\tau_{bh}}{\tau_\mathrm{total}}$ \text{rounded down to value in $\mathcal{C}$}
\State \text{Append $B_b$ to $\mathcal{B}$}

\For{$j=1,...,E$}

\State $B_j = B_F \frac{\tau_{jh}}{\tau_\mathrm{total}}$ \text{rounded down to value in $\mathcal{C}$}
\State \text{Append $B_j$ to $\mathcal{B}$}
\EndFor

\State $\ell_h = \frac{1}{N}\frac{1}{E}\sum_n (\tau_{bh} \ell_b + \sum_j \tau_{jh} \ell_j(T_n) )$ \text{ weighted centroid}

\State \textbf{Return:} $\ell_h, \mathcal{B}$
\end{algorithmic}

\label{alg:heur}
\end{algorithm}

\vspace{-0.5cm}

\subsection{Conventional Simulated Annealing}
\label{sec:sa}
SA~\cite{SA-survey} is a well-known probabilistic meta-heuristic technique that requires the definition of problem-specific parameters and functions in order to be efficient. The overview of the Conventional SA (Conv. SA) method is described in Alg.~\ref{alg:sa}.

\textbf{SA parameters: }The inputs are the trajectories $\mathcal{L}$ and minimum rate requirements $\tau^{\mathrm{min}}$ of the FENs (Line 1). The initial candidate solution is given by Conv. H, described previously (Line 3). The initial SA temperature ($t=t_{\mathrm{max}}$) is cooled down over $s_{\mathrm{max}}$ iterations, following a linear cooling schedule (Lines 2, 14, 15). The Maximum Temperature value $t_{\mathrm{max}}$ is empirically determined to achieve and 80\% target acceptance rate for neighbour solutions under a constant cooling schedule~\cite{SA-survey}. Linear cooling was selected for its balance between performance and computational efficiency in addressing the current problem\footnote{Future work can include the study of other cooling schedule options such as exponential or logarithmic.}. The temperature value decrease is related to the decreasing probability of accepting an objectively worse neighbour solution (i.e., negative $\Delta U$), throughout the search process (Lines 6-10). The fitness of a solution is evaluated by calculating the utility value $U$ of Eq.~(\ref{eq:problem1}), i.e., the weighted average capacity (Line 7).

\begin{algorithm}
\caption{Conventional SA Method}
\begin{algorithmic}[1]
\State \textbf{Input:} $\mathcal{L}, \tau^{\mathrm{min}}$ \textbf{Output:} $\ell_h^\star, \mathcal{B}^\star$

\State {Initialize:} $t = t_{\mathrm{max}}, s = 0,$
\State $(\ell_h, \mathcal{B}) \leftarrow $\text{ compute \textit{Heuristic} (Alg. 1)} with $(\mathcal{L}, \tau^{\mathrm{min}})$
\State \text{Update} $(\ell_h^\star, \mathcal{B}^\star) \leftarrow (\ell_h, \mathcal{B})$
\While{$s < s_{\mathrm{max}}$}

\State $(\ell_h^n, \mathcal{B}^n) \leftarrow $\text{ compute \textit{Neighbour}} with $(\ell_h, \mathcal{B})$
\State $\Delta U = U(\ell_h^n, \mathcal{B}^n) - U(\ell_h, \mathcal{B}) $

\If{$random(0,1) < e^{\frac{\Delta U}{t}}$} 

\State \text{Update} $(\ell_h, \mathcal{B}) \leftarrow (\ell_h^n, \mathcal{B}^n)$
\EndIf

\If{$U(\ell_h^n, \mathcal{B}^n) > U(\ell_h, \mathcal{B})$} 
\State \text{Update} $(\ell_h^\star, \mathcal{B}^\star) \leftarrow (\ell_h^n, \mathcal{B}^n)$

\EndIf
\State $s \leftarrow s + 1$
\State $t \leftarrow t_{\mathrm{max}}(s-s_{\mathrm{max}})/s_{\mathrm{max}}$

\EndWhile

\State \textbf{Return:} $(\ell_h^\star, \mathcal{B}^\star)$

\end{algorithmic}
\label{alg:sa}
\end{algorithm}

\textbf{Proposed Neighbour Function:} The problem has a high dimensional space that grows with the number of links in the network and the time periods. Exploring the search space efficiently is challenging, as there is no universal approach for generating neighbors. We design a neighbour function that aims to produce small $\Delta U$ changes. First, the \textbf{neighbour HAP position} is obtained using $\ell_h^n = \ell_h + \Delta \ell$ where $\ell_h$ is the current HAP position solution and $\Delta \ell$ is the small change introduced. We consider a random value for $\Delta \ell = rand(.) \times \Delta z$ where the random function produces a value in the interval $[-1, 1]$ with $\Delta z$ being the grid granularity step size in meters. For the \textbf{neighbour channel bandwidth assignment} $\mathcal{B}^n$ a list with all possible channel bandwidth assignment combinations is generated. From this list, the relevant combinations are filtered in two steps. First, we remove all combinations whose total bandwidth usage goes over the $B_F$ limit. On the other hand, we remove the combinations that are under-utilizing the available spectrum, considering a fixed threshold $\beta B_F$, where $\beta=0.8$ is empirically defined. After this setup, for each new $\mathcal{B}^n$ requested, we randomly select one from the pruned list, and evaluate if there is an improved utility value $U$; otherwise, the current assignment $\mathcal{B}$ is kept.

\subsection{Proposed SAFnet Method}

\label{sec:proposedmethod}

This section describes the Proposed SA for predictive Flying Networks (Prop. SAFnet) method, which incorporates penalty functions into an enhanced utility function. The constraints imposed by these functions reduce the number of valid solutions, making the search process more challenging. In demanding scenarios, such as those with wider coverage zones or higher traffic demands, it may be impossible to find a solution that meets all constraints. If all randomly generated neighbour solutions are discarded for violating constraints, the initial solution provided by Conv. H cannot be improved. 
However, in real-world applications, the algorithm must still provide a feasible solution. To address this, a penalty factor $p_f$ is introduced to account for constraint violations, incrementing its value each time a violation occurs and thereby impacting the solution's fitness, as detailed in Alg.~\ref{alg:uti}. Next, we specify how these constraints are handled and how they affect the utility value of each candidate solution.

\begin{algorithm}
\caption{Proposed SAFnet Method}
\begin{algorithmic}[1]
\State \textbf{Input:} $\mathcal{L}, \tau^{\mathrm{min}}$ \textbf{Output:} $\ell_h^\star, \mathcal{B}^\star$

\State {Initialize:} $t = t_{\mathrm{max}}, s = 0,$
\State $(\ell_h, \mathcal{B}) \leftarrow $\text{ compute \textit{Heuristic} (Alg. 1)} with $(\mathcal{L}, \tau^{\mathrm{min}})$
\State \text{Update} $(\ell_h^\star, \mathcal{B}^\star) \leftarrow (\ell_h, \mathcal{B})$

\While{$s < s_{\mathrm{max}}$}

\State {Initialize:} $u=0$, $p_f=0$
\State $(\ell_h^n, \mathcal{B}^n) \leftarrow $\text{ compute \textit{Neighbour}} with $(\ell_h, \mathcal{B})$

\State \text{Update} $p_f \leftarrow p_f + C^d$ \text{with $(\ell_h^n,\mathcal{L}$)}
\State \text{Update} $p_f \leftarrow p_f + C^{BW}$ \text{with $(\mathcal{B}^n,B_F$)}
\For{$n=1,...,N$}
\State \text{Calculate} $\tau_{bh}(T_n)$ \text{using (3) with $(\ell_h^n, \mathcal{B}^n)$}
\For{$j=1,...,E$}
\State \text{Calculate} $\tau_{jh}(T_n)$ 
\text{using (3) with $(\ell_h^n, \mathcal{B}^n)$}
\State \text{Update} $p_f \leftarrow p_f + C^{\mathrm{min}}(\tau_{jh}(T_n), \tau_{jh}^{\mathrm{min}})$
\State \text{Update} $u \leftarrow u + w_{j} \tau_{jh}(T_n) /N$
\EndFor
\State \text{Update} $p_f \leftarrow p_f + C^{bkh}(\tau_{bh}(T_n), \sum^E_j \tau_{jh}(T_n))$
\EndFor
\State $U(\ell_h^n, \mathcal{B}^n) = u - p_f \times u$

\State $\Delta U = U(\ell_h^n, \mathcal{B}^n) - U(\ell_h, \mathcal{B}) $

\If{$random(0,1) < e^{\frac{\Delta U}{t}}$} 

\State \text{Update} $(\ell_h, \mathcal{B}) \leftarrow (\ell_h^n, \mathcal{B}^n)$
\EndIf

\If{$U(\ell_h^n, \mathcal{B}^n) > U(\ell_h, \mathcal{B})$} 
\State \text{Update} $(\ell_h^\star, \mathcal{B}^\star) \leftarrow (\ell_h^n, \mathcal{B}^n)$

\EndIf
\State $s \leftarrow s + 1$
\State $t \leftarrow t_{\mathrm{max}}(s-s_{\mathrm{max}})/s_{\mathrm{max}}$

\EndWhile

\State \textbf{Return:} $(\ell_h^\star, \mathcal{B}^\star)$

\end{algorithmic}
\label{alg:uti}

\end{algorithm}

The \textbf{HAP Location Constraints} are given by Eq.~(\ref{eq:c-location}) and Eq.~(\ref{eq:c-samepos}). The candidate solutions are filtered during the selection process, considering the coverage zone constraint. $\ell_h$ is proposed using a stochastic selection process, and a repairing method is used whenever an infeasible solution is proposed, modifying it to become feasible (i.e., adjust the candidate $\ell_h$ solution to the closest position inside the coverage zone). For the minimum distance constraint, we set $C^d=1$ if $d^{\mathrm{min}} \geq d_{jh}(T_n)$ for any $n$ period; otherwise, $C^d$ is set to $0$. 
The \textbf{Budget Bandwidth Constraint} given by Eq.~(\ref{eq:c-channel}) is controlled with the binary parameter $C^{BW}$. The proposed neighbour Function explained in Sec.~\ref{sec:sa} automatically discards any $\mathcal{B}$ configuration that violates this constraint. However, this is not necessarily true for other benchmark algorithms, such as discrete exhaustive search. Specifically, $C^{BW}=1$, if $B_{F} \leq \sum_{B_{ih}\in \mathcal{B}} B_{ih}$; $C^{BW}=0$, otherwise.
The \textbf{QoS Constraints} are given by Eq.~(\ref{eq:c-min}) and Eq.~(\ref{eq:c-bkh}). Unlike the previous constraints, they are evaluated for each time period and each FEN. We set $C^{\mathrm{min}}=\frac{1}{NE}$, if the minimum rate requirement $\tau_{jh}^{\mathrm{min}} \leq \tau_{jh}(T_n)$ is not satisfied, incrementing the penalty factor, and $C^{\mathrm{min}}=0$ otherwise, thereby assigning a penalization that is proportional to the frequency of outage occurrences. Similarly to $C^{\mathrm{min}}$, $C^{bkh}=\frac{1}{N}$, if $\tau_{bh}(T_n) \leq \sum_{j}^E\tau_{jh}(T_n)$, i.e., the backhaul link capacity is insufficient to accommodate the sum-rate over all FEN links, and $C^{bkh}=0$ otherwise. With the normalization of the penalty by the time periods and number of FEN, the severity of the QoS constraint violation is addressed. 
When no constraints are violated, $p_f$ is set to 0 and $U$ is equal to the weighted average rate over the FEN links, as defined by the objective function in Eq.~(\ref{eq:problem1}).

\section{Performance Evaluation}

\subsection{Simulation Settings}

The network area consists of a square with dimensions $Z=(500,500)$ meters. For the SA-based methods, several parameters were empirically defined for the simulations, including $t_{\mathrm{max}}=10^{8}$, $s_{\mathrm{max}}=10^{4}$ and $\Delta z=5$. Table~\ref{tab:simulation-param} shows other fixed parameters used for the studied scenarios. 

\begin{table}
    \centering
    \caption{Simulation Parameters}
    \vspace{-0.2 cm}

    \begin{tabular}{lc|l}
         Description & Symbol & value\\
         \hline
         Time period duration (s) & $\Delta T$ & 0.1 \\
         Transmit Power (dBm) & $P^T$ & 20 \\
         Wavelength (mm) & $\lambda$ & 60 \\
         Power Spectral Density (dBm/Hz) & $N_0$ & -174 \\
         Supported Channel Bandwidth (MHz) & $\mathcal{C}$ & \{20, 40, 80, 160\}\\
         Frequency Band bandwidth (MHz) & $B_F$ & 320 \\ 
         Minimum Node Distance (m) & $d^{\mathrm{min}}$ & 1 \\ 
    \end{tabular}
    \vspace{-0.6 cm}
    \label{tab:simulation-param}
\end{table}

A different random seed is generated per simulation, where each seed impacts the random initial position of the FENs in the network, their trajectories and the distribution of weights and minimum QoS requirements. Despite the randomness on the definition of these variables, we control the process so that the position of the FENs are never outside the coverage zone, and the weights $w_j$ are sampled randomly from the set $\{w_j \in \mathbb{N}~|~w_j \leq 5,\}, \forall j \in \{1,...,E\}$, being normalized afterwards. 
Prop. SAFnet is evaluated against the benchmarks defined in Sec.~\ref{sec:refmethods} for an increasing number of FENs and total minimum rate requirements. Additionally, the Discrete Exhaustive Search (DES) method is considered, which selects a solution with the highest utility value $U$ defined in Eq.~(\ref{eq:problem1}) among an extensive discrete set of solutions. The coverage zone is discretized as a grid with edge length $\Delta z=1$m  and every combination of channel bandwidth assignment is attempted. DES does not scale for a large number of links or smaller $\Delta z$, as its computation becomes intractable due to the exponential growth in candidate solutions. However, it provides insight into the approximate optimality gap of other methods in small scale scenarios.
The evaluated network performance metrics are average throughput, outage probability and time complexity. The average capacity of the FEN links is given by $\frac{1}{N}\sum_{n}^N \sum_{j}^E \tau_{jh}(T_n)$ and their outage probability is given by $\frac{1}{NE} \sum_{n}^N \sum_{j}^E I\{\tau_{jh}(T_n) \geq \tau^{\mathrm{min}}_{jh}\}$, where the indicator $I$ is $1$ when the condition is met and $0$ otherwise.

\subsection{Time Complexity Analysis}

When comparing the complexity of benchmarks, DES method has time complexity $\mathcal{O}\left(NE\left(\frac{Z}{\Delta z}\right)^3 |\mathcal{C}|^{E}\right)$, where $|\mathcal{C}|$ is the number of valid channel bandwidth configuration values. Conv. H has the lowest time complexity $\mathcal{O}(NE)$ and Prop. SAFnet and Conv. SA have the same time complexity given by $\mathcal{O}(NEs_{\mathrm{max}})$. The common factor in the complexity analysis of each of the considered algorithms is due to the utility function value, which has a time complexity $\mathcal{O}(NE)$.

\subsection{Simulation Results}

\begin{figure}
    \centering
    \includegraphics[width=.75\linewidth]{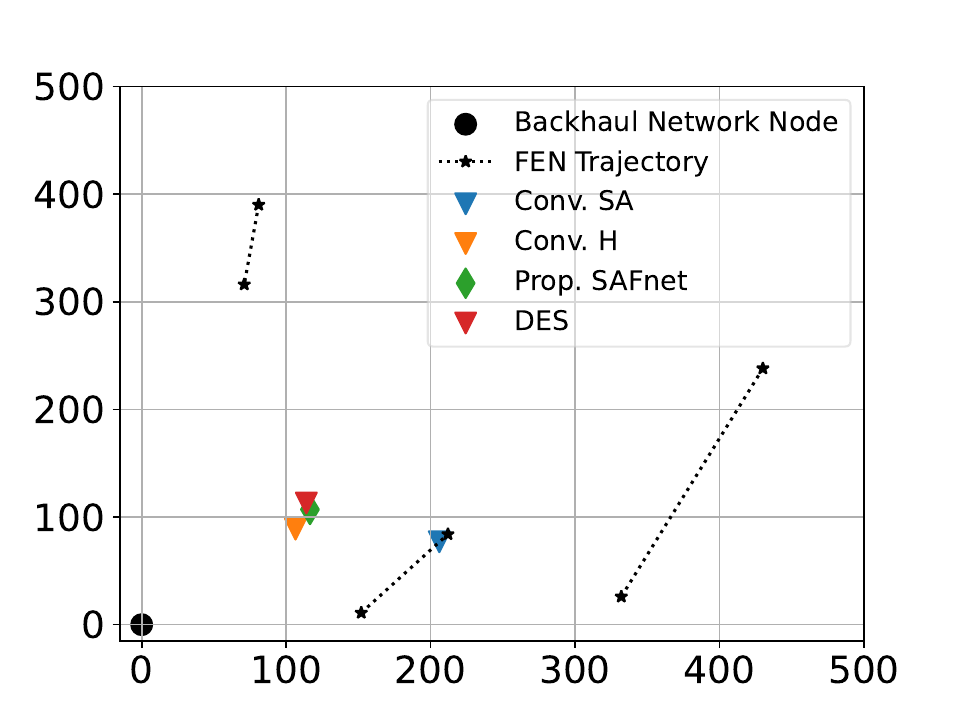}
    \vspace{-0.5 cm}
    \caption{HAP positions defined by the benchmark algorithms and Prop. SAFnet.}
    \label{fig:toy_topology}
    \vspace{-0.5 cm}

\end{figure}

\begin{figure*}
  \centering
    \subfloat[$E$ impact on Avg. FEN Link Outage.\label{fig:fen-fen-out}]{
      \includegraphics[trim={0 0cm 0 1cm}, clip, width=.34\textwidth]{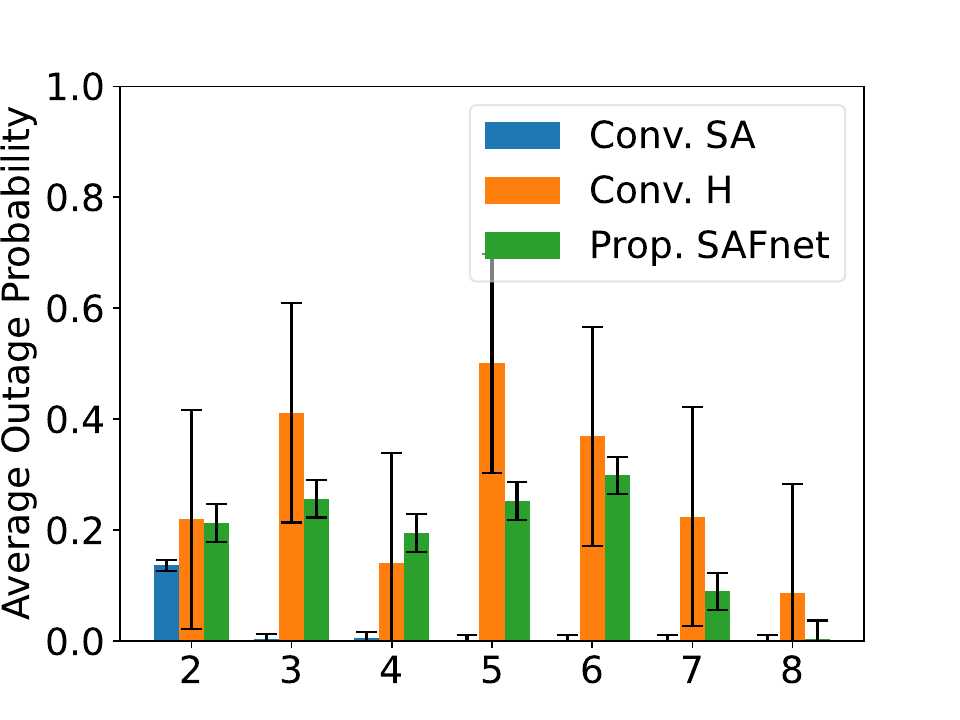}   
  }
  \hspace{-0.8 cm}
  \subfloat[Total rate requirements (bit/s) impact on Avg. FEN Link Outage.\label{fig:fen-min-out}]{
      \includegraphics[trim={0 0cm 0 1cm}, clip, width=.34\textwidth]{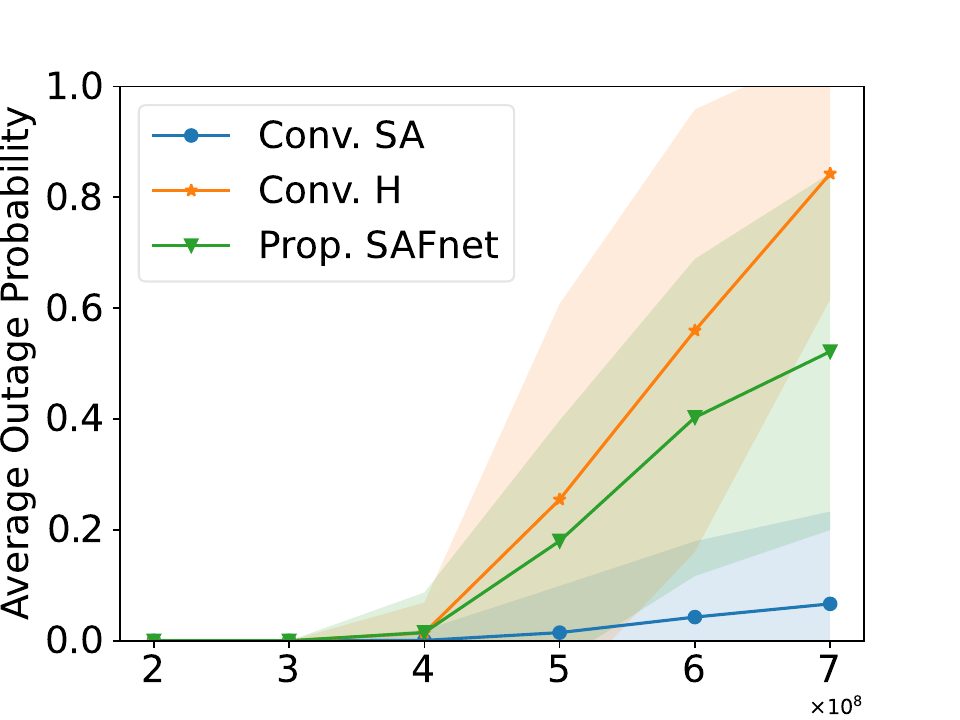}
  }
  \hspace{-0.8 cm}
  \subfloat[Average FEN Link Throughput.\label{fig:fen-cap}]{
      \includegraphics[trim={0 0cm 0 1cm}, clip, width=.34\textwidth]{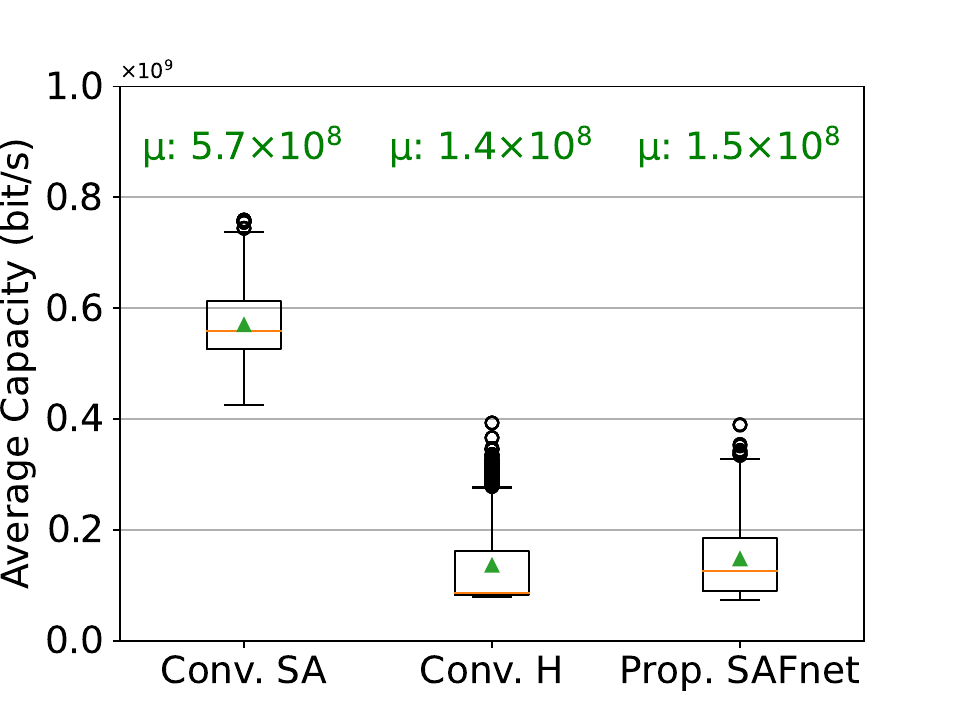}
  }
  \vspace{-0.0 cm}
  \subfloat[$E$ impact on Backhaul Link Outage.\label{fig:bkh-fen-out}]{
    \includegraphics[trim={0 0cm 0 1cm}, clip, width=.34\textwidth]{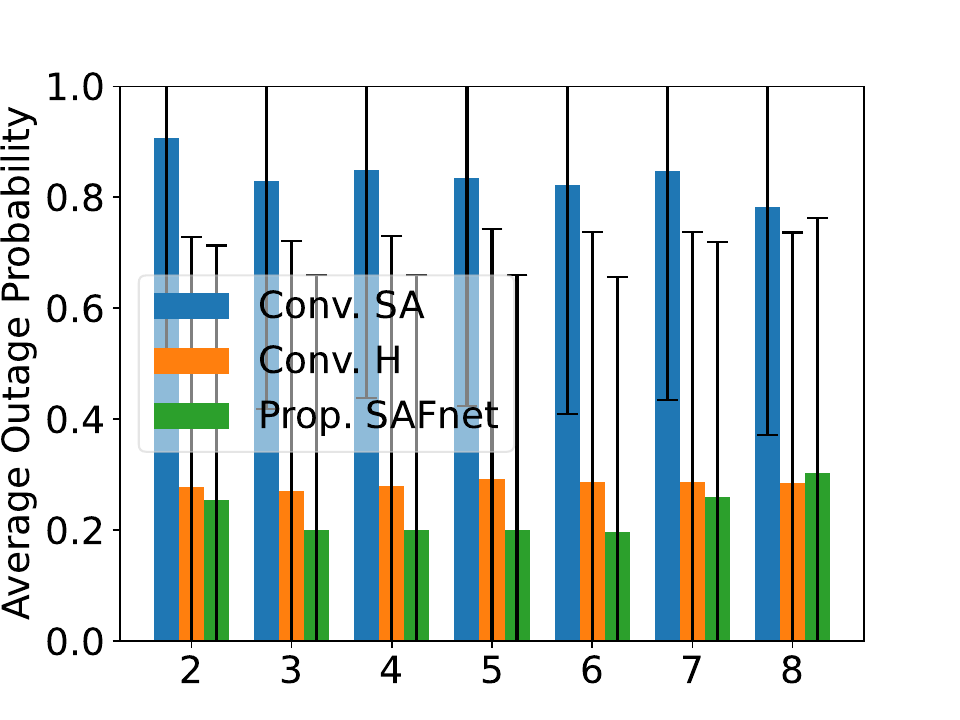}  
  }  
  \hspace{-0.8 cm}
  \subfloat[Total rate requirements (bit/s) impact on Backhaul Link Outage.\label{fig:bkh-min-out}]{
      \includegraphics[trim={0 0cm 0 1cm}, clip, width=.34\textwidth]{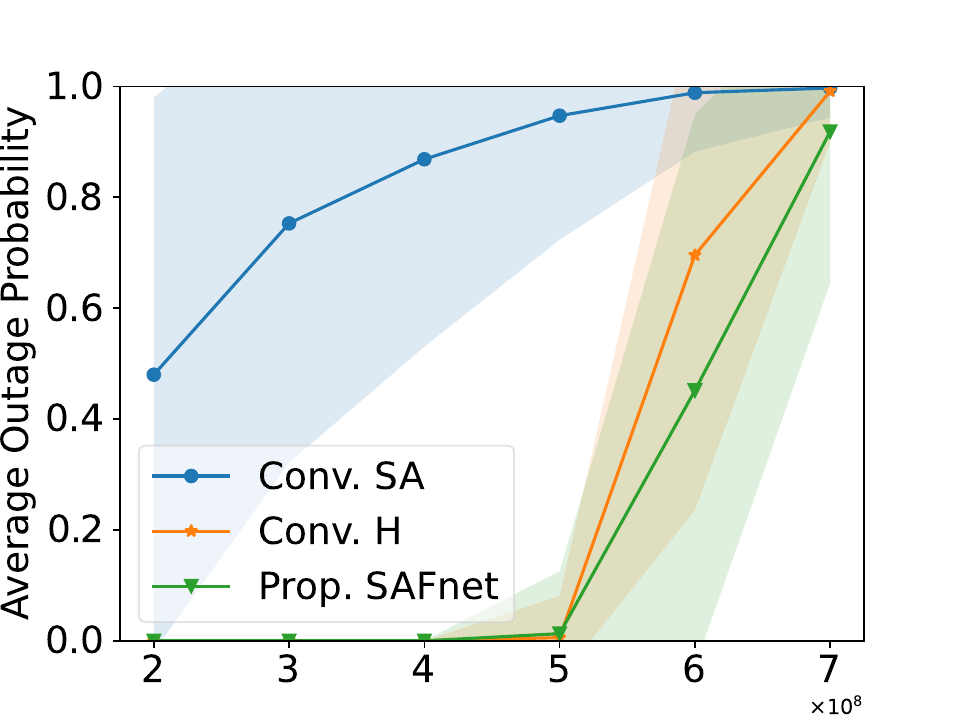}
  }
  \hspace{-0.8 cm}
  \subfloat[Backhaul Link Throughput.\label{fig:bkh-cap}]{
      \includegraphics[trim={0 0cm 0 1cm}, clip, width=.34\textwidth]{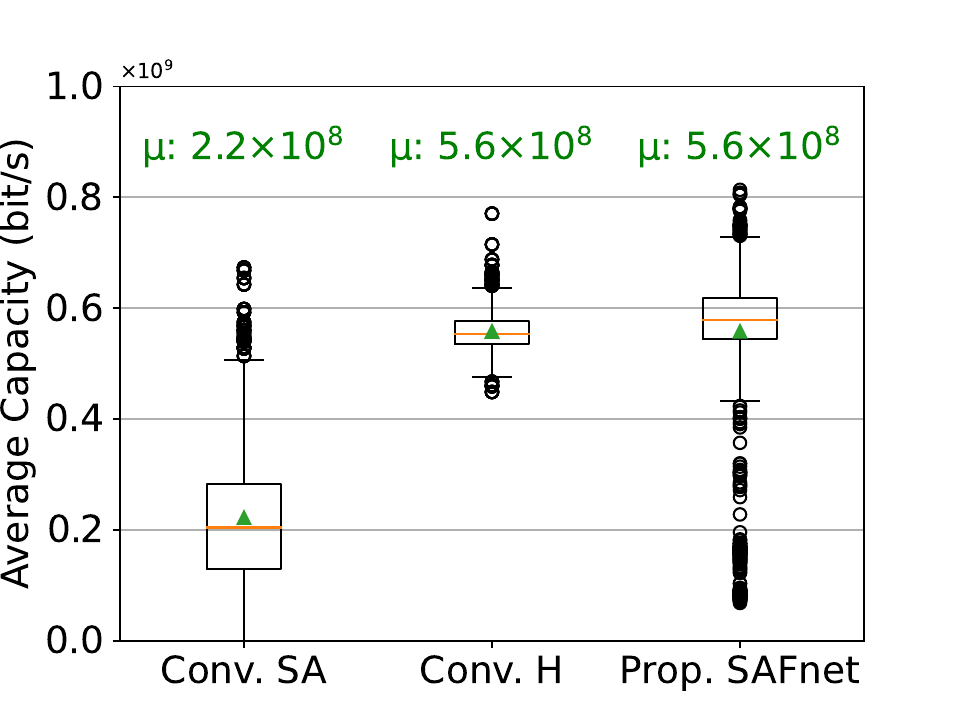}
  }
  \caption{Network performance metrics for larger network scenarios with varying number of FENs and rate requirements.}
  \vspace{-0.6 cm}
  \label{fig:ext}
\end{figure*}

The first scenario consists of a single backhaul node and 3 FENs with random trajectories, throughout 30 simulated seconds. All 3 FENs share the same weight value ($w_j=0.33$) as well as minimum rate requirements ($\tau^{\mathrm{min}}_{jh}=150$ Mbits/s). Fig.~\ref{fig:toy_topology} shows the coverage zone of the scenario as well as the corresponding HAP position given by each of the benchmarks, including DES. Note that Conv. H, as the initial candidate solution of Conv. SA and Prop. SAFnet, is already positioning it at a relatively close distance to DES, showing that it is a reasonable initial solution with significantly lower complexity. Conv. SA, which does not consider the existing constraints, positions the HAP where the fitness function is maximized, closer to the FENs, disregarding the backhaul link. 

Regarding the bandwidth assignment component of each solution, Prop. SAFnet converged to the same assignment strategy as DES, with $\mathcal{B}=\{80,40,40,160\}$, while Conv. H computed a sub-optimal assignment $\mathcal{B}=\{40,40,40,160\}$, not fully using the available frequency bandwidth. On the other hand, Conv. SA, which is not penalized for going over budget, converged to $\mathcal{B}=\{160,160,160,20\}$. This results in a permanent outage of the backhaul link, as shown in Table~\ref{tab:toy-results}. Conv. H method achieves higher backhaul throughput, at the cost of the average FEN throughput, which results in a 4\% outage probability. Both DES and Prop. SAFnet balance the link throughput better, with zero outage probability. Moreover, Prop. SAFnet has a slight gain in backhaul throughput, due to the discretization granularity used in DES ($\Delta z=1$m). 

\begin{table}
\centering
\caption{First Scenario: Network Performance Results}
    \vspace{-0.2 cm}

\begin{tabular}{l|cccc}
            & \multicolumn{1}{c}{\begin{tabular}[c]{@{}c@{}}Backhaul  \\ Thp. (bit/s)\end{tabular}} 
            & \multicolumn{1}{c}{\begin{tabular}[c]{@{}c@{}}Avg. FEN \\ Thp. (bit/s)\end{tabular}} 
            & \multicolumn{1}{c}{\begin{tabular}[c]{@{}c@{}}Backhaul  \\ Outage\end{tabular}}  
            & \multicolumn{1}{c}{\begin{tabular}[c]{@{}c@{}}Avg. FEN  \\ Outage\end{tabular}}  \\ \hline
DES & 5.78$\times 10^{8}$ & 2.03$\times 10^{8}$ & 0 & 0   \\ \hline
Conv. H & 6.01$\times 10^{8}$ & 1.61$\times 10^{8}$ & 0 & 0.04   \\ \hline
Conv. SA & 8.43$\times 10^{7}$ & 5.82$\times 10^{8}$  & 1 & 0   \\ \hline
Prop. SAFnet & 5.81$\times 10^{8}$ & 2.03$\times 10^{8}$ & 0 & 0  
\end{tabular}
\label{tab:toy-results}
\vspace{-0.8 cm}

\end{table}

After validating the performance of the benchmarks and the Prop. SAFnet, which is comparable to that of the DES approach in the first scenario, we now present the extensive simulation results under the parameters in Table~\ref{tab:simulation-param}. These simulations aggregate 100 different random seeds, for each combination of number of FENs $E=\{2,3,...,8\}$ and rate requirements $\sum \tau_{jh}^{\mathrm{min}}=\{2\times 10^{8},3\times 10^{8},...,7\times 10^{8}\}$ bit/s. These results exclude the DES method due to the exponential increase in combinations as $E$ grows. Figs. \ref{fig:ext}\subref{fig:fen-fen-out}, \ref{fig:ext}\subref{fig:fen-min-out} show the average FEN link outage probabilities of the benchmarks against an increasing number of FENs and sum-rate of all requirements $\tau^{\mathrm{min}}_{jh}$, respectively. Similarly, Figs. \ref{fig:ext}\subref{fig:bkh-fen-out}, \ref{fig:ext}\subref{fig:bkh-min-out} evaluate the backhaul link. The colored regions and whiskers are given by $\mu \pm \sigma$ with $\mu$ and $\sigma$ representing the sample mean and standard deviation of the outage probability distribution, respectively. Note that when studying the impact of increasing $E$, we are averaging the results of all the combinations of $\tau^{\mathrm{min}}_{jh}$, and 
when studying the impact of increasing $\tau^{\mathrm{min}}_{jh}$, we are averaging the results of all the combinations of $E$.
Conv. SA has the lowest average FEN link outage but the highest backhaul link outage, regardless of the parameters we are studying, showing the limitations of this approach and importance of the proposed penalty functions. The performance of Conv. H is best in the case of an even number of FENs, as seen when $E=4$, and worse for the odd numbers. This is related to the sub-optimal bandwidth assignment problem identified in the first scenario. When comparing the mean outage results for the average FEN link and backhaul link, SAFnet has a 32\% and 18\% improvement respectively, relatively to the best benchmark (Conv. H).
Figs. \ref{fig:ext}\subref{fig:fen-cap}, \ref{fig:ext}\subref{fig:bkh-cap} show the average capacity obtained for both links. The first and third quartile form the box with a line that represents the median. The whiskers extend the box by $1.5\times$ the interquartile range of the box, with outliers being all the samples outside the whiskers. We aggregate the results of all numbers of FENs and range of rate requirements resulting in a highly diverse set of simulation scenarios, justifying the outliers numbers. 
Although Conv. SA has the highest link capacity, its high backhaul outage rate renders it impractical. Comparing SAFnet and Conv. H, which offer a better balance of outage probability, SAFnet achieves a 7\% capacity advantage over Conv. H on average FEN link.

In summary, the results clearly show that Conv. SA fails to deliver valid solutions, making Conv. H the most reliable benchmark among the methods tested. However, Prop. SAFnet significantly outperforms Conv. H reducing backhaul link outages by 18\%, improving average FEN link outage probability by 32\%, and increasing the link capacity by 7\%. Although Prop. SAFnet has a higher time complexity than Conv. H, it remains far more efficient than DES, making it a superior choice for scenarios where computational resources, time constraints, and QoS requirements are critical. Ultimately, the choice between Conv. H and Prop. SAFnet may depend on the specific demands and constraints of the mission at hand.  

\vspace{-0.1cm}
\section{Conclusions and Future Work}
\vspace{-0.1cm}

Predictive FNs, where UAVs operate along predefined missions and trajectories, present unique challenges in optimizing bandwidth allocation and relay positioning. We proposed SAFnet, a low-complexity SA-based approach tailored for these scenarios. By incorporating the proposed neighbour and penalty functions, SAFnet leverages prior knowledge of UAV trajectories to significantly enhance network performance. Our simulations demonstrate that SAFnet reduces relay link outage probability by 32\% and backhaul link outage by 18\% while boosting overall network throughput by 7\% on average, compared to the best benchmark.

In future work, we aim to evolve SAFnet to address real-world practical constraints, including shared frequency resources, inter-node interference, and scalability with a larger number of HAPs and backhaul nodes.

\bibliographystyle{IEEEtran}
\vspace{-0.1cm}
\bibliography{refs}

\end{document}